\documentclass[aps,prb,showpacs,twocolumn,superscriptaddress]{revtex4-1}

\usepackage{graphicx}
\usepackage{amsmath,amssymb}
\usepackage{hyperref}
\usepackage{natbib}
\usepackage{bm}

\begin{document}


\title{Topological properties of the bond-modulated honeycomb lattice}

\date{\today}
\author{F.~Grandi}
\affiliation{Dipartimento di Scienze Fisiche, Informatiche e Matematiche, Universit\`a di Modena e Reggio Emilia, Via Campi 213/A, I-41125 Modena, Italy }
\affiliation{Scuola Internazionale Superiore di Studi Avanzati (SISSA), Via Bonomea 265, I-34136 Trieste, Italy}
\author{F.~Manghi}
\affiliation{Dipartimento di Scienze Fisiche, Informatiche e Matematiche, Universit\`a di Modena e Reggio Emilia, Via Campi 213/A, I-41125 Modena, Italy }
\affiliation{Centro S3, CNR - Istituto Nanoscienze, Via Campi 213/A, I-41125 Modena, Italy}

\author{O.~Corradini}
\affiliation{Facultad de Ciencias en F\'isica y Matem\'aticas, Universidad Aut\'onoma de Chiapas, Ciudad Universitaria, Tuxtla Guti\'errez 29050, M\'exico}
\affiliation{Dipartimento di Scienze Fisiche, Informatiche e Matematiche, Universit\`a di Modena e Reggio Emilia, Via Campi 213/A, I-41125 Modena, Italy }
\author{C.M.~Bertoni}
\affiliation{Dipartimento di Scienze Fisiche, Informatiche e Matematiche, Universit\`a di Modena e Reggio Emilia, Via Campi 213/A, I-41125 Modena, Italy }
\affiliation{Centro S3, CNR - Istituto Nanoscienze, Via Campi 213/A, I-41125 Modena, Italy}
\begin{abstract}
 We study the combined effects of lattice deformation, {\it e-e} interaction and spin-orbit coupling in a two-dimensional (2D) honeycomb lattice. We adopt  different kinds of hopping modulation---generalized dimerization and a Kekul\'e distortion---and calculate topological invariants for the non-interacting system and  for the interacting system. We identify  the parameter range (Hubbard $U$, hopping modulation, spin-orbit coupling) where the 2D system behaves as a  trivial insulator or   Quantum Spin Hall Insulator.

\end{abstract}

\pacs{71.10.Fd, 71.27.+a, 73.43.-f, 73.22.-f }
\maketitle

\section{Introduction}
The study of novel topological  phases of matter is an extremely active research field~\cite{RevModPhys.82.3045, Ando, RevModPhys.83.1057}. Quantum Spin Hall Insulators  (QSHI) are  a remarkable example of topology at work; in these two-dimensional insulating  systems   the non-dissipative spin current carried by gapless  edge states owes its robustness to the topology of bulk bands  described by the non-zero value of the $\mathbb{Z}_{2}$ topological invariant.
The two-dimensional graphene-like lattice with  intrinsic spin-orbit coupling has been identified as a paradigmatic example of QSHI~\cite{PhysRevLett.95.146802,PhysRevB.74.195312,PhysRevB.76.045302}.
The spin-orbit helical interaction, described by a  nearest-neighbor spin-dependent complex hopping, opens a gap in the otherwise linear spectrum of the honeycomb lattice and at the same time induces a metallic behavior on the edges.

A band gap opening,   the \emph{conditio sine qua non} for the emergence of topological  features, can be achieved---at least conceptually---in different ways, not all of them with the same  topological consequences.
A gapped phase on the honeycomb lattice may be induced by modulating the tight-binding hopping amplitudes to describe different kinds of bond dimerization or by including many-body {\em e-e} interactions.
The interplay between these three types of ``gapping" interactions---spin-orbit couplings, hopping modulation and on-site {\em e-e} interaction---has been recently studied assuming the  bond dimerization that can be  associated to uniaxial strain~\cite{PhysRevB.87.205101}. Another interesting hopping modulation is the one leading to a Kekul\'e  distortion where stronger and weaker  nearest-neighbor links alternate on the honeycomb lattice  in a $\surd 3 \times \surd 3 $ arrangement. This structure turns out to be stable in the presence of
nearest-neighbor and next-to-nearest-neighbor {\em e-e} interactions~\cite{PhysRevB.81.085105,PhysRevLett.107.106402} resulting, at the mean field level, in an effective bond dimerization of a  Kekul\'e type.

In this work, we explore the combined effects of spin-orbit couplings, hopping modulation and on-site {\em e-e} interaction. We superimpose different kinds of hopping modulation on a  Kane-Mele-Hubbard model~\cite{PhysRevB.82.075106,PhysRevB.84.205121,PhysRevLett.106.100403,PhysRevB.85.115132} describing  a two-dimensional honeycomb lattice in the presence of both  spin-orbit coupling and local {\em e-e} interaction and we identify  the topological properties of this interacting system  in terms of topological invariants. This will be done by solving the many-body problem within Cluster Perturbation Theory (CPT) and extracting topological invariants from the many-body Green's function.  The goal is to clarify how the topological phases that stem from the intrinsic  spin-orbit coupling are modified by different kinds of hopping texturing and by {\em e-e} interaction.

The paper is organized as follows: in section \ref{sec1} we consider non-interacting electrons in the presence of intrinsic spin-orbit coupling and  different kinds of hopping modulation; section \ref{sec2} extends the results to the interacting case, and the last section is devoted to discussion and conclusions.

\section{Single-particle description}
\label{sec1}

In the present section we consider a honeycomb model of non-interacting electrons represented by a single particle hamiltonian with a spin-dependent hopping term
\begin{equation}
\label{kmh}
\hat{H} = \sum_{i l, i' l' s} t_{il,i'l'}(s)\hat{c}_{i l s}^{\dag} \hat{c}_{i' l'  s}~.
\end{equation}
Here $i,i'$ run over the atomic positions within the unit cell, $l,l'$ refer to lattice vectors identifying  the unit cells of  the lattice, and {\em s\,}=1,2  is for spin up and down. $\hat c^\dagger_{ils}$ and $\hat c_{ils}$ are respectively the electron creation and electron annihilation operators. The hopping term  $t_{il,i'l'}(s)$  includes both the  first-neighbor spin-independent hopping  and the Haldane-Kane-Mele
second-neighbor spin-orbit coupling~\cite{PhysRevLett.61.2015,PhysRevLett.95.146802} given by
$
\imath t_{KM} s_{z} ({d_{1}} \times {d_{2}})_{z}
$, where $d_{1}$ and $d_2$ are  unit vectors  along the two bonds that connect site  $i l$ with site $i' l'$,  and $s_z$ is the unit vector in the direction orthogonal to the lattice plane.
This hamiltonian  preserves time-reversal symmetry and parity symmetry. In turn, this implies that---by Kramer's
theorem---states come in time-reversal pairs and the Chern number computed from the bulk occupied bands
identically vanishes, and so does the charge conductivity. However the spin-conductivity may be non-vanishing as it depends upon the difference between the two spin-filtered Chern numbers~\cite{PhysRevLett.95.146802}.

\begin{figure}[h]
 \centering \includegraphics[width=8.5cm]{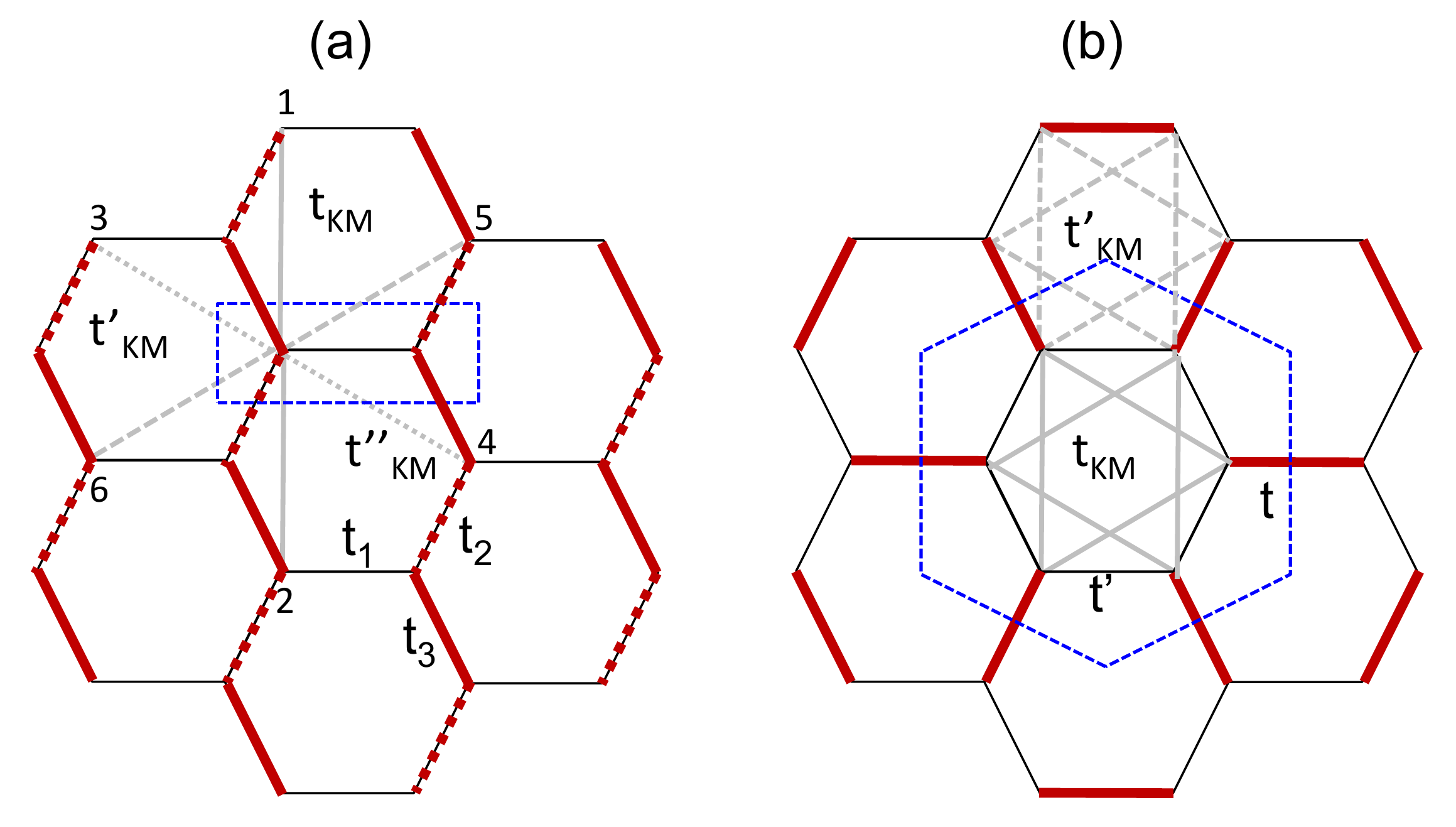}
  \caption{\label{geo} (Color online) Geometry of the two-dimensional (2D) honeycomb lattice with different hopping texture: (a) generalized dimerization with different nearest-neighbor hopping parameters; (b) Kekul\'e distortion. The unit cells  in the two cases, containing  two and six atoms respectively, are also shown.  Different values of second neighbor interactions are indicated.}
\end{figure}

We start by considering the modulation in the hopping amplitudes among nearest-neighboring sites that may arise as a consequence of a non-uniform shear strain. As shown in Ref.~\onlinecite{PhysRevB.80.045401} this corresponds to different values for the three nearest neighbor hopping parameters (Fig. \ref{geo} (a)).
Since the system has time-reversal and inversion symmetry we may identify the topological character of the system through the $\mathbb{Z}_2$ parity invariant  defined as the exponent $\Delta$ in the expression \cite{PhysRevB.76.045302}
\begin{align}
\label{z2}
(-1)^{\Delta} =\prod_{TRIM} \prod_{n=1}^{N} \eta_{n}(\Gamma_i)~.
\end{align}
where $\eta_{n}(\Gamma_i)= \pm 1 $ are the parity eigenvalues  of the occupied bands for any of the two spin sectors, calculated  at time-reversal invariant momenta (TRIM), and $\Gamma_i$ is defined by the condition that $-\Gamma_i= \Gamma_i +\mathcal{G}$ with $\mathcal{G}$ a reciprocal lattice vector.
The value of the $\mathbb{Z}_2$ topological invariant  distinguishes trivial insulators ($\Delta =0$, mod 2) from topological QSH insulators ($\Delta =1$, mod 2).

The eigenvalues and eigenvectors for the honeycomb lattice as well as the parity eigenvalues at TRIM points can be easily calculated analytically in terms  and hopping parameters by solving the $2\times 2$ secular problem; we obtain
\begin{eqnarray*}
  \eta_{1}(\Gamma_1) &=& 1 \\
  \eta_{1}(\Gamma_2) &=& {\rm sign}[t_1+(t_2-t_3)] \\
  \eta_{1}(\Gamma_3) &=& {\rm sign}[t_1-(t_2-t_3)] \\
  \eta_{1}(\Gamma_4) &=& {\rm sign}[t_1-(t_2+t_3)]
\end{eqnarray*}
where the $t_i$ are the three (generically different) hopping parameters. For the  $\mathbb{Z}_2$  invariant we thus have
\begin{equation}
\label{inv_gg}
(-1)^{\Delta} = {\rm sign} \left[ t_{1} - \left( t_{2} + t_{3} \right) \right] {\rm sign} \left[ t_{1}^{2} - \left( t_{2} - t_{3} \right)^{2} \right]
\end{equation}

\begin{center}
\begin{figure}[bth]
 \centering \includegraphics[width=17cm]{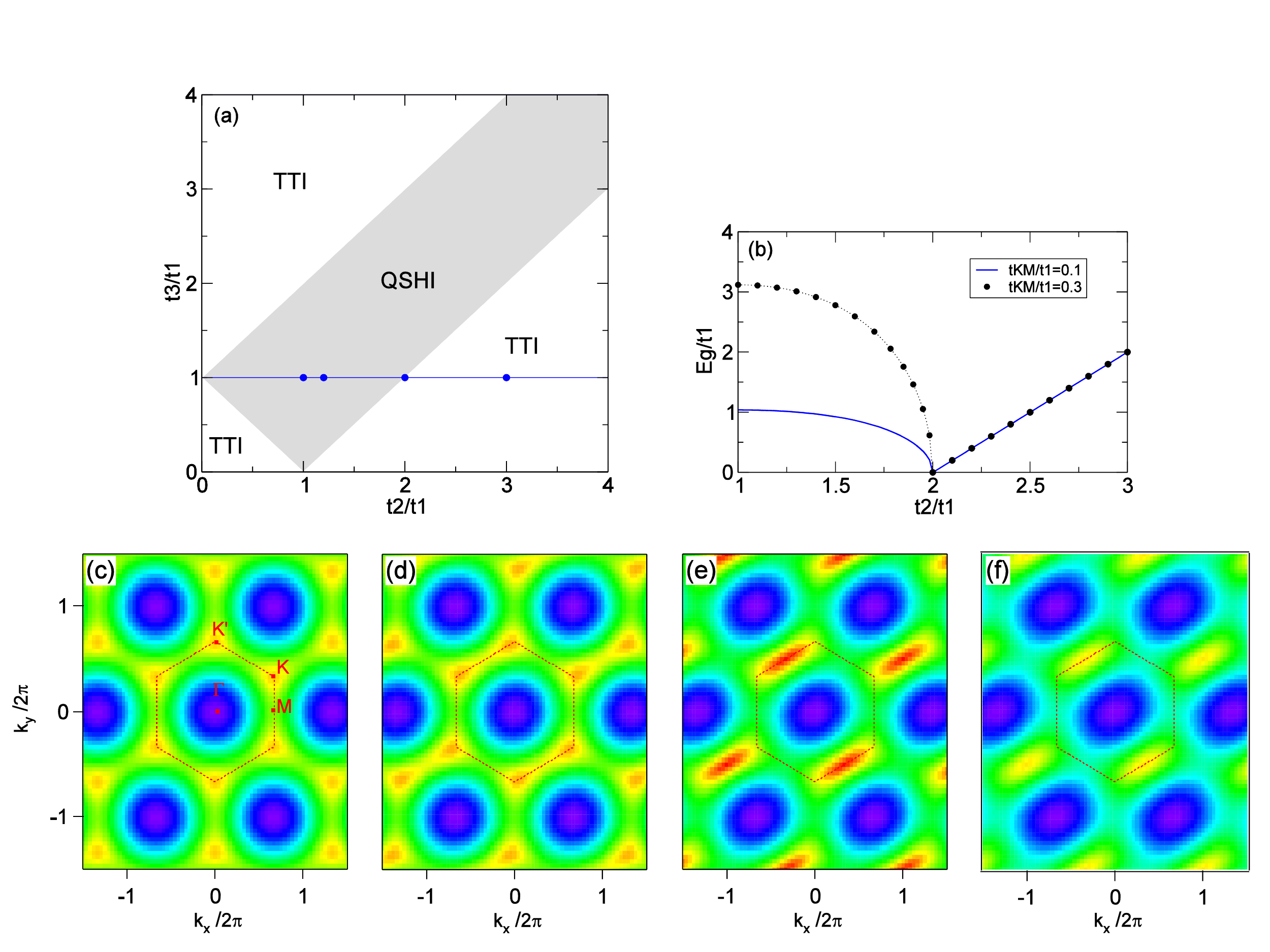}
  \caption{\label{dimer_nonint} (Color online) Upper panel: (a) Phase diagram for Kane-Mele model of the honeycomb lattice assuming the  generalized bond dimerization of Fig. 1 (a). The two phases, QSHI and TTI, correspond to different values of the $\mathbb{Z}_2$  invariant ($\Delta=1$ and $ \Delta=0$ respectively, see text).  (b) Gap value as a function of the hopping parameter $t_2$  assuming $t_{1} = t_{3}=1, t_{KM}=t'_{KM}=t''_{KM} $. Continuous line  is for $t_{KM}/t_1=0.1$, dotted line for $t_{KM}/t_1=0.3$. Lower panel:  Density plots of the occupied energy states as a function of k-point with $t_{KM}=0.1$. The evolution of the band structure is considered for $t_{KM}=t'_{KM}=t''_{KM} $ in a subset of first nearest neighbor hopping parameters with $t_{1} = t_{3}=1 $ indicated as blue dots in panel (a): $t_{2}=1 $ (c), $t_{2}=1.2 $ (d),  $t_{2}=2 $ (e) , $t_{2}=3 (f) $. In the color scale, red indicates zero gap. Brillouin zones (thin red dotted lines) and high symmetry points are also shown.}
\end{figure}
\end{center}

Interestingly, these quantities depend just on first near- est neighbor hopping parameters and not on second nearest neighbor hoppings describing spin-orbit interaction.
This is a direct consequence of symmetry: each site in the unit cell is connected to 3 pairs of second nearest neighbors sharing the same distance but placed in opposite directions (sites 1-2, 3-4, 5-6 in Fig. \ref{geo}). For this reason the Hamiltonian matrix $H(k)$ at any TRIM  point  does not depend on  Haldane-Kane-Mele second-neighbor spin-orbit coupling. We have in fact
\begin{equation*}
H(k) = \left(
\begin{array}{cc}
g(k) & -f(k)  \\
-f(k)^* &  - g(k)
\end{array}
\right)
\end{equation*}
where
\begin{eqnarray*}
g(k) &=& 2[t_{KM} \sin(k_y ) \\  \nonumber
     &-& t'_{KM} \sin(k_x \sqrt{3}/2+k_y /2)  \\  \nonumber
     &-&  t''_{KM} \sin(- k_x\sqrt{3}/2+ k_y /2)] \\  \nonumber
f(k) &=&  t_1 \exp(\imath k_x / \sqrt{3}) \\  \nonumber
     &+& \exp[-\imath k_x /(2 \sqrt{3}) ]  [t_2  \exp(- \imath  k_y / 2)  + t_3  \exp(\imath k_y / 2)].
\end{eqnarray*}
It is then easy to check that $g(k)$ at any TRIM  point is identically zero.
Still, the spin-orbit interaction is essential in order to obtain a non-trivial topological behavior: for $t_{KM}= 0$ a modulation in the hopping parameters would only transform a semi-metal into a trivial insulator while for $t_{KM}\neq 0$ the system is  always an insulator  (except just at the phase boundary---see below). For $t_{KM}\neq 0$, by tuning the hopping parameters $t_i$  we may go from a trivial insulating phase to a topological Quantum Spin Hall insulating regime. The values where a phase transition occurs---either between QSHI and topologically trivial insulator (TTI) for $t_{KM}\neq 0$ or between semimetal and TTI for $t_{KM}= 0$---are identical and the phase diagram that we obtain  (Fig.~\ref{dimer_nonint} (a))  coincides, as far as the phase separations are concerned, with the one reported in refs.~\onlinecite{PhysRevB.80.045401,PhysRevB.74.033413} for $t_{KM}= 0$. We observe that right at the transition  between QSHI and TTI phases the single particle  gap $\Delta _{sp}$ closes down  and the system recovers a semi-metallic behavior. We analyze in particular the behavior of the system varying just one hopping parameter ($t_2$) with $t_{KM}=t'_{KM}=t''_{KM}$,  moving in the parameter space along the line shown  in Fig.~\ref{dimer_nonint} (a) where  $t_3=t_1 $. For this choice of parameters the gap between filled and empty states evolves as shown in Fig.~\ref{dimer_nonint} (b). Before the transition, the absolute value of the energy gap depends on $t_{KM}$ but after the transition it becomes independent on $t_{KM}$ and  increases linearly as $ E_{g} = 2 \left( t_{2} - 2 t_{1} \right)$.

In the lower panel of Fig.~\ref{dimer_nonint} we show   the evolution of the band structure assuming $t_{KM}=t'_{KM}=t''_{KM}$ and   $t_{KM}/t_1= 0.1$ along the line in parameter space  where $t_{1} = t_{3} $. By increasing the value of $t_{2}$ the positions in k-space of the  band gap  move along a line parallel to $k_{y} = \frac{1}{\sqrt{3}} k_{x}$ and merge at M points where the gap closes down for $t_{2}/ t_{1}=2$, signaling the topological transition from QSHI to TTI. We stress again that  the spin-orbit term, even including a modulation in second-neighbor hopping interaction,   does not alter the  parity symmetry ($C_2$) and as such it does not affect the gap position in k-space.  After the transition the gap remains at M.

We turn now to the Kekul\'e distortion. In this case  the unit cell contains six atoms, with alternating values in  the nearest-neighbor hopping parameters $t $ and $t'$  as shown in Fig. \ref{geo} (b). In principle we   have  two possible values of the second nearest-neighbor parameters; since their dependence on the lattice deformation is not easy to assess we have considered two separate cases, namely $t'_{KM}=t_{KM}$ and $t'_{KM}=t'/t \times t_{KM}$ as suggested in Ref.~\onlinecite{PhysRevB.85.205102}.

\begin{center}
\begin{figure}[bth]
 \centering \includegraphics[width=17cm]{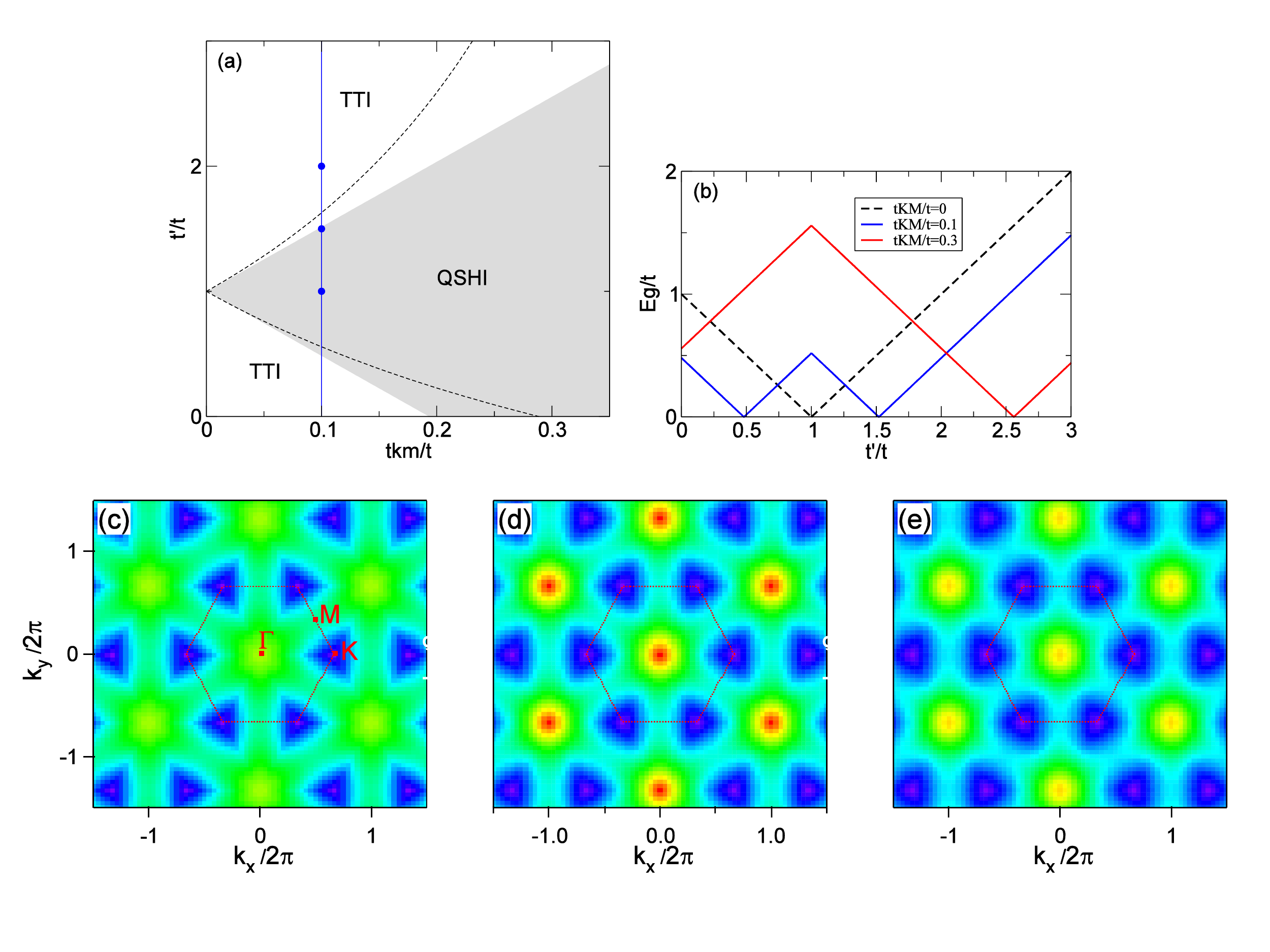}
  \caption{\label{kekule0}  (Color online) (a) Phase diagram for the 2D honeycomb lattice with Kekul\'e distortion showing the topological behavior  as a function of nearest neighbor modulation $t'$ and spin-orbit coupling $t_{KM}$. The grey area corresponds to $t'_{KM}=t_{KM}$ while the dashed line is the phase separation for $t'_{KM}= t'/t \times t_{KM}$.  (b) Gap value as a function of the hopping parameter $t'$  for different values of $t'_{KM}=t_{KM}$ as indicated in the inset.  Lower panel: Density plots of the occupied energy states as a function of k-point with $t'_{KM}=t_{KM}=0.1$. The evolution of the band structure is considered in a subset of first nearest neighbor hopping parameters ( $t'/t=1 $ (c), $t'/t=1.5 $ (d),  $t'/t=2 $ (e)) indicated as blue dots in panel (a). In the color scale, red indicates zero gap. The reduced Brillouine zone for the 6-site unit cell (thin red dotted line) and high symmetry points are shown.}
\end{figure}
\end{center}

Neglecting at first the variation in the second nearest-neighbor hopping parameters we notice that the minimum separation between filled and empty states  is, for any value of $t'/t$ and $t_{KM}/t$, pinned at a $\Gamma$ point. This is  shown in the lower panel of Fig. \ref{kekule0} where the dispersion of the highest occupied band is shown for different values of $t'/t$ and for $t_{KM}/t=0.1$. This remains true also for  $t'_{KM} \neq t_{KM}$. We may then identify  the analytic dependence of the $\mathbb{Z}_{2}$ invariant on the hopping parameters by considering the eigenvalues at the $\Gamma$ point only, where the diagonalization of the $6 \times 6 $ Hamiltonian matrix is trivial, and look for the conditions that guarantee a zero gap between filled and empty states. Indeed  in the non-interacting case the transition from a trivial to a non-trivial topological phase requires the gap to close down.
The six eigenvalues at $\Gamma$ ($e_1=-t-2t'$, $e_2=t+2 t'$, $e_3=t-t'-  \sqrt{3} (t_{KM}+2 t'_{KM})$, $e_4=-t+t'- \sqrt{3} (t_{KM}+2 t'_{KM})$, $e_5=t-t'+ \sqrt{3} (t_{KM}+2 t'_{KM})$, $e_6=-t+t'+ \sqrt{3} (t_{KM}+2 t'_{KM})$) are easily obtained from the hamiltonian matrix
\begin{equation}
H_{\uparrow} (\Gamma) = \left(
\begin{array}{cccccc}
0 & -t' & m  & -t & m^* & -t' \\
-t' & 0 & -t' & m  & -t & m^* \\
m^* & -t' & 0 & -t' & m  & -t \\
-t & m^* & -t' & 0 & -t' & m \\
m & -t & m^* & -t' & 0 & -t' \\
-t' & m & -t & m^* & -t' & 0
\end{array}
\right)
\end{equation}
where $m=\imath (2 t'_{KM}+t_{KM})$. Therefore the gap closure occurs when one of the following two conditions is verified
\begin{equation}
\begin{split}
& t-t'- \sqrt{3} (t_{KM} +2 t'_{KM})= 0 \\
& t-t'+\sqrt{3} (t_{KM} +2 t'_{KM}) = 0
\end{split}
\end{equation}
This leads  to the following analytic expression for the $\mathbb{Z}_{2}$ invariant as a function of the hopping parameters:
\begin{equation}
\label{inv_kek}
(-1)^{\Delta} = {\rm sign} \left[ \left(t-t' \right)^{2} - 3 \left( t_{KM}+2 t'_{KM} \right)^{2} \right]~.
\end{equation}
This relation has been checked numerically in terms of  parity eigenvalues according to eq.~\eqref{z2}. We obtain in this way the  phase  diagram of Fig. \ref{kekule0} (a): any $t_{KM}\neq0$ defines a range of    $t'/t $ where the system behaves as a QSHI, and this range depends on the strength of the spin-orbit coupling. The hopping modulation has different effects if $t_{KM}=0$ or $t_{KM}\neq 0$  as shown in Fig. \ref{kekule0} (b) where the evolution of the gap value is reported  for $t_{KM}=t'_{KM}$: for the undistorted system in particular we have  zero gap and  maximum gap for $t_{KM}=0$ and $t_{KM}\neq 0$, respectively. Notice that  the analytic expression of the $\mathbb{Z}_{2}$ invariant allows us to obtain quite simply the phase diagram also in the case of $t'_{KM}\neq t_{KM}$ and that for $t'_{KM}=t'/t \times t_{KM}$  we recover the results~\footnote{Notice that  for $t'_{KM}=t'/t \times t_{KM}$  we plot  $1/\alpha$ vs $\lambda$, with $\alpha \equiv t / t'$, $\lambda\equiv t_{KM}/t = t'_{KM}/t'$  while  in Ref. \onlinecite{PhysRevB.85.205102} the corresponding phase diagram is  $\alpha$ vs $\lambda$ } of Ref.~\onlinecite{PhysRevB.85.205102}.

\section{Effects of  {\em e-e} correlation}
\label{sec2}
The Kane-Mele-Hubbard model \begin{align}
\label{kmh_int}
\hat{H} = \sum_{i l, i' l' s} t_{il,i'l'}(s)\hat{c}_{i l s}^{\dag} \hat{c}_{i' l'  s} +
U \sum_{i l} \hat{c}_{i l \uparrow}^{\dag}\hat{c}_{i l \uparrow}\hat{c}_{i l \downarrow}^{ \dag}\hat{c}_{i l  \downarrow}~,
\end{align}
where  on-site {\em e-e} repulsion is added to the non-interacting hamiltonian of eq.~\eqref{kmh}, is a paradigmatic example of an interacting topological insulator~\cite{Assaadrev,doi:10.1142/S0217984914300014}.  In this case, in order to topologically characterize the system, we face two distinct problems: on one side we need to substitute the single particle band structure with the quasi-particle excitation energies that can be obtained from the many-body Green's function; on the other side we must  extend the $\mathbb{Z}_{2}$ parity invariant, originally associated to a single particle state, to the interacting case. In Refs.~\onlinecite{PhysRevB.85.165126,PhysRevX.2.031008} it has been demonstrated that the $\mathbb{Z}_{2}$  invariant is determined by the behavior of the one-particle propagator at zero frequency only:
the inverse of the Green's function at zero frequency defines a fictitious  noninteracting topological hamiltonian \cite{0953-8984-25-15-155601}
\begin{align} h_{topo}(k)\equiv -G^{-1}(k,0)
\label{htopo}
\end{align}
and its eigenvectors
\begin{align}
\label{eigen}
h_{topo}(k)|k,n\rangle = \epsilon_{n }(k)|k,n\rangle
\end{align}
are the quantities that in eq.~\eqref{z2} replace the non-interacting band eigenvectors to obtain the topological invariant for the interacting system.

These concepts have been recently applied to identify the topological character of heavy fermion mixed valence compounds~\cite{PhysRevB.88.035113,PhysRevLett.111.176404,PhysRevLett.110.096401,PhysRevB.87.085134} and of the half-filled honeycomb lattice~\cite{grandi} also in the presence of uniaxial  bond dimerization~\cite{PhysRevB.87.205101}.

In order to solve the eigenvalue problem~\eqref{eigen}, in strict analogy with what is done in any standard Tight-Binding scheme for non-interacting hamiltonians, a Bloch basis  expression of the topological hamiltonian, namely, of the dressed Green's function and of its inverse, is required:
\begin{align}
\label{gij}
 G_{ij}(k, \omega)=\langle \Psi_0|\hat{c}_{k  i }^{\dag}\hat{G} \hat{c}_{k j}|\Psi_0 \rangle
 + \langle \Psi_0|\hat{c}_{k  i }\hat{G} \hat{c}_{k j}^{\dag}|\Psi_0\rangle
\end{align}
where $\hat{G}=\frac{1}{\omega-\hat{H}}$
and
\begin{align*}
\hat{c}_{k i}^{\dag}=\frac{1}{\sqrt{L}}\sum_{l}^{L} e^{-i k\cdot(R_l+r_i)}\hat{c}_{l i }^{\dag} \, ; \ \ \hat{c}_{k i}=\frac{1}{\sqrt{L}}\sum_{l}^{L} e^{ik\cdot(R_l+r_i)}\hat{c}_{l i }
\end{align*}
with $R_l$ the  lattice vectors ($L$ $\rightarrow \infty$) and $r_i$ the atomic positions inside the unit cell.

\onecolumngrid
\twocolumngrid

Here we calculate the dressed Green's function by  Cluster Perturbation Theory  (CPT) \cite{Senechal}. CPT   belongs to the class of Quantum Cluster theories\cite{RevModPhysQC} that solve the problem of many interacting electrons in an extended lattice by a  \emph{divide-and-conquer } strategy, namely by solving first the many body problem in a subsystem of finite size and then embedding it within  the infinite medium. Different Quantum Cluster approaches (Dynamical Cluster Approach \cite{DCA}, Cellular Dynamical Mean Field Theory  \cite{Senechal_CDMFT,PhysRevLett.110.096402}, and Variational Cluster Approaches \cite{VCA}) differ for the embedding procedure and/or for the way the lattice Green's function---or the corresponding self-energy---is expressed in terms of the cluster one. The common starting point is  the choice of the $M$-site cluster used to \emph{tile} the extended lattice.
By construction CPT is exact in the two limits
$U/t=0$ (non-interacting band limit), $ U/t =\infty$ (atomic limit); for intermediate values of $U/t$ it opens a gap in metallic systems at half occupation \cite{ManghiMnO}.

\onecolumngrid
\twocolumngrid

In CPT,  Green's function (\ref{gij}) for the extended lattice is calculated by solving the equation
\begin{equation}\label{QC}
G_{i j}(k,   \omega)=G^c_{ i j}(  \omega)+ \sum_{i'}^M B_{i i'}(k,  \omega) G_{i' j}(k,   \omega).
\end{equation}
Here $G^c_{ i j}$ is the cluster Green's function in the local basis obtained by exact diagonalization of the interacting hamiltonian for the finite cluster; we separately solve the   problem for  $N$, $N-1$, and $N+1$ electrons and
express the cluster Green's function in the Lehmann representation at real frequencies.

The matrix $B_{i i'}(k,\omega)$ is given by
\begin{align*}
B_{i i'}(k,\omega)= \sum_l^L e^{i k\cdot R_l} \sum_{i''}^M G^c_{i i''}(\omega) t_{i'' 0, i' l}(s)
\end{align*}
where $t_{i''0 ,i' l}$ is the hopping  between site $i'$ and $i''$ belonging to different clusters.

The key approximation here is the expression of  the complete Green's function in terms of Green's functions of decoupled clusters and it is important to verify the accuracy of the results by using larger and larger cluster sizes. This procedure is limited by the dimensions of Hilbert space  used in the exact diagonalization, dimensions that grow exponentially with the number of  sites. A further limitation in the cluster choice arises by a symmetry requirement since only clusters that preserve the point group symmetries of the lattice must  be used.~\cite{RevModPhysQC,grandi} The role of symmetry in Quantum Cluster approaches is complex:  the extended system is described as a periodic repetition  of correlated units and   the translation periodicity  is preserved only  at the superlattice level. In the  honeycomb lattice where the Dirac cones are the consequence of  perfect long-range order,  theories based on Quantum Cluster schemes, such as CDMFT, VCA, and CPT, regardless of them being variational or not, and independent on the details of the specific implementations (different impurity solvers, different temperatures),   at $t_{KM}=0$  give rise to a spurious  excitation gap  for  $U \rightarrow 0$. The only exception is the Dynamical Cluster Approach  (DCA)  that  preserves by construction translation symmetry and has been shown to describe better the small $U$ regime; DCA becomes, however, less accurate at large $U$  where it overemphasizes the semimetallic  behavior of the honeycomb lattice.~\cite{PhysRevB.87.205127} In this sense DCA and the other Quantum Cluster approaches can be considered as complementary and it would be interesting to compare their results also for the distorted honeycomb lattice.
Another strategy has been proposed that seems to overcome this shortcoming,  providing for the undistorted honeycomb lattice  a semimetal behavior up to some finite $U$.~\cite{PhysRevB.90.165136} The strategy consists in  choosing clusters  that break  the lattice point $C_6$ symmetry (8- and 10-site clusters).  The quasiparticle band dispersion that is obtained in this way is, however, unphysical: quasiparticle energies at $k$ and $Rk$, $R$ being a point group rotation, turn out to be different,  violating a very basic rule of band structure.~\cite{grandi}  And it is just this violation that makes the system semimetallic at finite $U$ since the gap closes at a $k$-point but not at its rotated counterpart. For this reason  breaking the rotational symmetry is not an allowed strategy to correct the erroneous insulating phase.

We have checked the dependence of our results on the cluster size by comparing the case of the generalized dimerization results obtained for 2- and 8-site clusters;   we have verified that no significant changes occur in the spectral functions  and for this reason we report results obtained only for the smallest cluster size, namely, 2- and 6-site clusters for the  generalized bond dimerization  and Kekul\'e distortion, respectively. Notice that in this case the clusters used to ``tile" the infinite lattice are those shown in Fig. \ref{geo} and that $t_1$ ($t'$) describe intra-cluster hoppings for the two distortions (generalized dimerization and Kekul\'e,  respectively).
Eq.~\eqref{QC} is  solved  by an $M\times M$  matrix inversion at each $k$ and $\omega$.
A second $M\times M$ matrix inversion is needed to obtain the topological hamiltonian according to eq.~\eqref{htopo}. The topological hamiltonian is then diagonalized and its  eigenvectors are used for the calculation of    $\mathbb{Z}_2$ according to~\eqref{z2}.

It is worth recalling that  the eigenvalues of $h_{topo}$ used to calculate the value of the $\mathbb{Z}_2$  invariant in principle have nothing to do with the quasi-particle excitation energies: they only  contain topological information and the full Green's function is needed to calculate quasi-particle spectral functions
\begin{align}
\label{akn}
A(k,\omega) = \frac{1}{\pi}\sum_n {\rm Im}\,  G(k,n,\omega)
\end{align}
where
 \begin{align*}
G(k,n,\omega) =  \frac{1}{M}\sum_{i i'} e^{-i k \cdot(r_i-r_{i'})}\alpha^{n*}_i(k) \alpha^n_{i'}(k)  G_{i i'}(k,\omega)
 \end{align*}
with $n$ the band index and $\alpha^n_i(k)$ the eigenstate coefficients obtained by the single-particle band calculation.~\cite{ManghiMnO}
Spectral functions   can also  be used to identify topological properties,  looking for the existence of gapless quasiparticle states in one-dimensional (1D) honeycomb ribbons.  The energy broadening necessarily  involved in the calculation of spectral functions makes this procedure  less accurate than the calculation of the ${\mathbb Z}_2$ invariant based on $h_{topo}$ eigenvectors: in $G^{-1}$ no energy broadening is required and  the boundaries in the phase diagram are sharply identified.

\begin{center}
\begin{figure}[tbh]
 \centering \includegraphics[width=18cm]{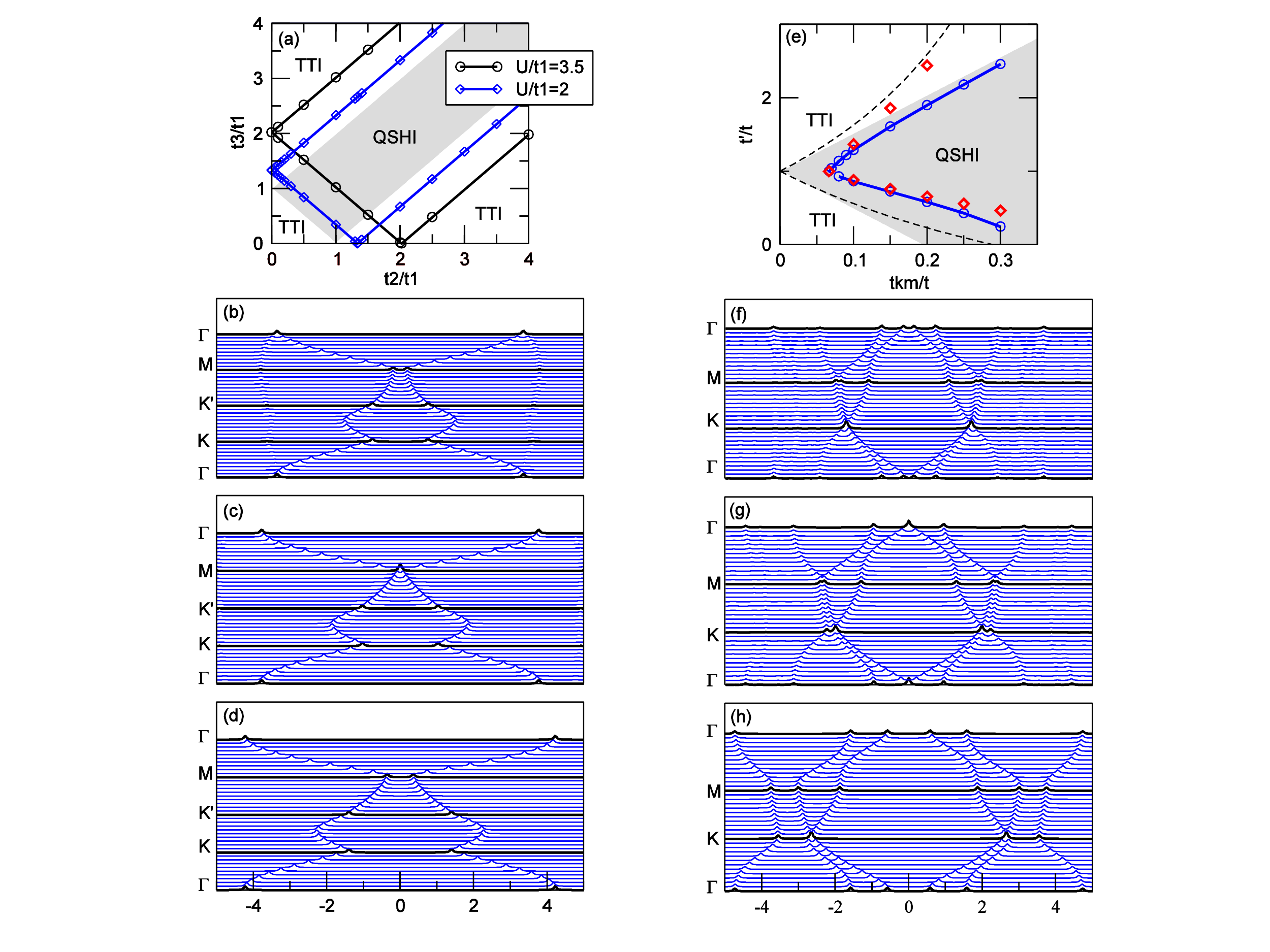}
  \caption{\label{akn_fasi_int} Phase diagram  and spectral functions for the Kane-Mele model of the honeycomb lattice in the presence of on-site {\em e-e} interaction for  the generalized bond dimerization (left) and for the Kekul\'e distortion (right panel). Panel (a): phase diagrams for the generalized bond dimerization obtained with different values of $U$  as a function of $t_2$ and $t_3$ at $t_{KM}=0.1$. Panels (b)--(d): spectral functions for the dimerized honeycomb lattice with  $t_{KM}=0.1$, $U=3$,  $t_2=t_3=1$, and (b) $t_1=1$, (c) $t_1=1.5$, and (d) $t_1=2$.
  Panel (e): phase diagram for the Kekul\'e distortion   at a fixed value $U=3$ as a function of $t'$ and $t_{KM}$.  The filled area corresponds in both cases to the non interacting result reported in Figs. \ref{dimer_nonint} and \ref{kekule0}. The dashed line  indicates the phase separation for for $t'_{KM}= t'/t \times t_{KM}$. In blue and red are reported the results assuming $t'_{KM}=t_{KM}$ and $t'_{KM}=t'/t \times t_{KM}$, respectively. Panels (f)--(h): spectral functions for the  honeycomb lattice in the Kekul\'e distortion with $U=3$,  $t=1$, $t'_{KM}=t_{KM}$, and (f) $t'=1$, (g) $t'=1.3$, and (h) $t'=2$.  }
\end{figure}
\end{center}

Fig. \ref{akn_fasi_int}  shows the results that  we obtain for  interacting electrons in the honeycomb lattice  with the two kinds of hopping modulation (generalized dimerization and Kekul\'e distortion).     By comparing the interacting case and  the non-interacting one we notice that  the QSHI/TTI phases are modified by the local {\em e-e} interaction for both lattice distortions. In the  case of generalized dimerization   (Fig. \ref{akn_fasi_int} (a))  the overall region in parameter space where the system is in the QSHI phase  increases with $U$ but at the cost of larger distortions: when on the contrary the system is almost undistorted (lower corner on the left of Fig. \ref{akn_fasi_int} (a), where $t_2$ and $t_3$ are closer to $t_1$) the effect of {\em e-e} is to extend the region of the TTI phase. For $U\geq 3.5$ the undistorted system is always topologically trivial.

The phase separation lines remain linear and independent on the strength of spin-orbit coupling, as in the non interacting case. Indeed, the effect of {\em e-e} interaction is to induce a  renormalization of the intra-cluster hopping parameters and therefore the topological hamiltonian of eq.~\eqref{htopo} coincides with an effective single-particle hamiltonian with modified hopping terms. This is particularly evident when a 2-site cluster is used as a basic unit in CPT, but remains true with larger clusters. We have checked this   by considering an 8-site cluster and we do not find significant differences.

In the Kekul\'e distortion, as expected from the previous analysis for the non-interacting case, different results are obtained assuming $t'_{KM}= t_{KM}$ or  $t'_{KM} \neq t_{KM}$. However in  both cases  the effect of {\em e-e} interaction favors  even more clearly the TTI phase since the total area where the system behaves as a QSHI is reduced with respect to the non-interacting case and, for a given distortion, larger values of spin-orbit coupling are required to have a non trivial topological character~\cite{Note1}.

The lower panels of Fig. \ref{akn_fasi_int}   show  the spectral functions  that we obtain for the two kinds of hopping modulation  as a function of the intra-cluster hopping parameters ($t_1$ and $t'$ respectively) at  fixed values of Hubbard $U$ and of inter-cluster hopping parameters $t'_{KM}=t_{KM}$ ($t_2=t_3=1$ and $t=1$ in the two cases respectively). We notice that the hopping modulation induces in both cases a closure of the energy separation between filled and empty quasi-particle states, signaling the topological phase transition.

\onecolumngrid
\twocolumngrid

\section{Conclusions}
\label{sec3}

We have studied the joint effects of intrinsic spin-orbit coupling, hopping modulation and on-site {\em e-e} interaction on the topological properties of the 2D honeycomb lattice. The goal was to understand how the topological phases induced by  intrinsic  spin-orbit coupling are modified by different kinds of lattice distortions and by {\em e-e} interaction.
The main results may be summarized as follows.
In the non-interacting case the shape of the phase diagram obtained assuming a generalized dimerization  does not depend on the value of the intrinsic spin-orbit coupling: the phase separation lines are identical for any value of $t_{KM}$,  with the noteworthy difference that in the parameter range where the system is for $t_{KM}=0$  a semi-metal, for   any other $t_{KM}\neq 0$  the system  is  a QSHI. In the absence of spin-orbit coupling the Kekul\'e distortion makes the system insulating and in this case the parameter space where the system behaves as a QSHI increases with  the value of $t_{KM}$.

We have extended the analysis in terms of topological invariants to the case of interacting electrons by calculating the dressed Green's function within CPT and the topological hamiltonian. For both lattice distortions, in the regime of relatively small deformations,  the effect of {\em e-e} interaction is to reduce the region where the system behaves as a QSHI.  In this sense we may conclude that  lattice distortions  and  {\em e-e} interaction do not cooperate in inducing a non trivial topological phase but rather reduce the possibility of finding the honeycomb lattice in a non-trivial topological state.


%
\end{document}